\documentclass{elsarticle}
\biboptions{authoryear}

\newcommand\arcdeg{$^{\circ}$}

\begin{document}

\begin{frontmatter}

  \title{Radio Observations of Magnetic Cataclysmic Variables}

  \author{Paul Barrett \corref{cor1}}
  \ead{pebarrett@gmail.com}
  \address{George Washington University, Washington DC, 20052, USA}
  \cortext[cor1]{Corresponding author}

  \author{Christopher Dieck}
  \address{United States Naval Observatory, Washington DC 20392, USA}

  \author{Anthony J. Beasley}
  \address{National Radio Astronomy Observatory, Charlottesville, VA
    22903, USA}

  \author{Paul A. Mason}
  \address{New Mexico State University, Las Cruces, NM, 88003, USA}
  \address{Picture Rocks Observatory, 1025 S. Solano, Suite D, Las
    Cruces, NM 88001, USA}

  \author{Kulinder P. Singh}
  \address{Indian Institute of Science Education and Research Mohali,
    Manauli, PO 140306, India}

  \begin{abstract}

    The NSF's Karl G. Jansky Very Large Array (VLA) is used to observe
    122 magnetic cataclysmic variables (MCVs) during three observing
    semesters (13B, 15A, and 18A).  We report radio detections of 33
    stars with fluxes in the range 6--8031 $\mu$Jy.  Twenty-eight
    stars are new radio sources, increasing the number of radio
    detected MCVs to more that 40. A surprising result is that about
    three-quarters (24 of 33 stars) of the detections show highly
    circularly polarized radio emission of short duration, which is
    characteristic of electron cyclotron maser emission. We argue that
    this emission originates from the lower corona of the donor star,
    and not from a region between the two stars. Maser emission
    enables a more direct estimate of the mean coronal magnetic field
    of the donor star, which we estimate to be 1--4 kG assuming a
    magnetic filling factor of 50\%.  A two-sample Kolmogorov-Smirnov
    test supports the conclusion that the distribution function of
    radio detected MCVs with orbital periods between 1.5--5 hours is
    similar to that of all MCVs.  This result implies that
    rapidly-rotating (P$_{spin} < 10$ days), fully convective stars
    can sustain strong magnetic dynamos. These results support the
    model of \citet{taam89} that the change in angular momentum loss
    across the fully convective boundary at P$_{orb} \approx 3$ hours
    is due to a change in the magnetic field structure of the donor
    star from a low-order to high-order multipolar field.
 
  \end{abstract}


\end{frontmatter}

\section{Introduction}

A project was initiated early in 2013 using the NRAO Very Long
Baseline Array (VLBA) to perform astrometry of six radio-bright ($>1$
mJy) magnetic cataclysmic variables (MCVs) (namely, BG CMi, AM Her, DQ
Her, ST LMi, GK Per, \& AR UMa) in an attempt to accurately measure
their distances. These observations were carried out in Semesters 13B
and 14A. Astrometry requires at least four observations at roughly
three month intervals with preferably two observations being at
quadrature for best results.  Unfortunately, source variability
resulted in less than four successful observations per star. As a
result, this project was only partially successful. The results of
these observations are in preparation. At the same time, a survey was
begun using the NSF's Karl G. Jansky Very Large Array (VLA) to
identify new radio-bright MCVs for follow-up VLBA astrometry. Previous
to the current study, only eight MCVs were known radio sources; four
polars (V834 Cen, AM Her, ST LMi, \& AR UMa) and four Intermediate
Polars (IPs; AE Aqr, BG CMi, DQ Her, \& GK Per). We note that
contemporaneous with this study, \citet{copp15,copp16} observed nine
nearby dwarf novae and nova-like CVs and detected eight of them at a
flux level of a few tens of $\mu$Jy. The success of the first VLA
survey in 2013 was motivation to expand this project to encompass all
known MCVs.

The present paper is a status report of this project as of July
2018. Section 2 is a summary of the observations and data analysis of
the surveys that have been performed since 2013, with particular
emphasis on those during VLA Semesters 13B, 15A and 18A. Section 3 is
a summary of the results of these three surveys, including the list of
detected sources. The observations and results of the complete survey,
including detections and non-detections, will be published in a
separate paper. In Section 4, we primarily focus on the circular
polarized emission from these sources. We discuss the most likely site
of this emission and its implication for the donor star's mean coronal
magnetic field and CV evolution. Our conclusions are given in Section
5.

\section{Observations and Data Analysis}

For our initial set of VLA observations, we observed 111 MCVs during
two observing semesters, 13B and 15A, at primarily three frequencies
(C-, X-, and K-bands; 4--6, 8--10, and 20--22 GHz, respectively) at
full polarization. A fourth frequency, Q-band (40--44 GHz), was also
used for a few observations during Semester 13B. No detections were
made at this frequency. For semester 13B, 40~hours of observing time
was requested to observe 60 of the optically ({\it B} magnitude)
brightest, and likely closest, MCVs from the \citet{ritt03}
\textit{Catalog of cataclysmic binaries, low-mass X-ray binaries and
  related objects (Seventh edition);} (sp., ed. 7.19) that are north
of declination $-40$\arcdeg. An equal number of targets were selected
from the polar and IP subclasses in order to avoid biasing the sample
toward the brighter IPs. The targets were also chosen to span a wide
range of local sidereal time. The observing program allowed about
three MCVs to be observed in each one hour scheduling block
(SB). Exposures were approximately two minutes per frequency. Each SB
was scheduled twice to increase the probability of detecting the
MCVs. The VLA scheduled 35 of the 40~hour, resulting in 42 MCVs being
observed. For semester 15A, 69~hour of observing time was requested to
observe another 69 MCVs in the \citeauthor{ritt03} catalog. Except for
one SB, each SB contained two MCVs with each exposure being
approximately five minutes per frequency. The VLA scheduled all
69~hour resulting in an additional 69 MCVs being observed. See
\citet{barr17} for a detailed discussion of the data analysis and
results of these observations.

We have also performed several additional surveys. For Semester 16B,
we observed AM Her and AR UMa for 21 hours at C-, X-, \& K-bands. An
additional 56 hours of time were used to observe another set of 56
MCVs during semesters 17B and 18A; 14 hours during Semester 17B, and
42 hours during Semester 18A. The latter two surveys were only
observed using the X-band, which enables longer ($\approx 20$ minutes)
exposures on each source. In summary, for the surveys to date, we have
obtained 195 out of 199 hours by performing 1 hour filler
observations. The data presented in this paper are from Semesters 13B,
15A, and 18A. The data from Semesters 13B and 15A are presented in
\citet{barr17}, whereas the data from Semester 18A are new. The
remaining observations will appear in a series of forthcoming papers.

As discussed in \citet{barr17}, all data are calibrated using the VLA
CASA automated calibration pipeline. The data from Semesters 13B and
15A use version 4.2.2, and Semester 18A uses version 5.1.2. The clean
(ver. 4.2.2) and tclean (ver 5.1.2) algorithms are then used for
source detection and flux measurements. Six polarization or Stokes
images (sp., I, Q, U, V, RR, and LL, where LL and RR are the left and
right polarization channels, resp.) are created for each observation
and target. The cleaning is performed in two steps: one for the I, Q,
U, and V planes; and one for the RR and LL planes. The Stokes I image
is usually best for detecting weakly polarized sources; while the
Stokes RR and LL images, for strongly polarized sources. (Henceforth,
circular polarization is implied when referring to polarization. No
sources show linear polarization.) The size of the image depends on
the VLA configuration and observing band. A typical image size for the
X band is $10 \times 10$ arcseconds.  The clean algorithm uses natural
weighting, because all sources are assumed to be point sources. The
cleaned (model) and residual images are used to measure the source
flux and the standard deviation of the noise, respectively. The
polarization and its error are calculated from the cleaned and
residual polarization images.

\section{Results}

\subsection{Radio Detections}

Table 1 is an abbreviated list of detections of 33 MCVS as of July
2018, i.e., from observing semesters 13B, 15A, and 18A.  A complete
list of detections and upper limits will be given in a forthcoming
paper.  Columns 1-4 are respectively the GVCS name, MCV subclass,
observing semester, and frequency band. The C, X, and K-band
frequencies for Semesters 13B and 15A are 4.464--6.512, 7.964--10.012,
and 20.060--22.108 GHz; and the X-band frequencies for Semesters 18A
are 7.928--12.024 GHz.  Columns 5--7 are the Stokes I, and the RR and
LL polarization fluxes in $\mu$Jy. The fluxes of the remaining Stokes
parameters (Q, U, \& V) are omitted. Column 8 is the percentage of
circular polarization. Column 9 is the total signal-to-noise (S/N) of
the detection. It is the product of the S/N of the Stokes I flux, the
circular polarization fluxes, and the PSF-normalized difference of the
observed and expected source positions. For highly polarized sources,
the S/N of the polarized flux can be much greater than the Stokes I
flux and therefore be the major factor in the total S/N. The algorithm
for the total S/N and the probability of source misidentification is
described by \citet{barr17}. Except for a few sources, most of the
observed positions are within a few tenths of an arcsecond of the Gaia
position \citep{gaia16, gaia18}. For sources having multiple
detections at a particular frequency, only the greatest flux is
listed. Because these radio sources are highly variable, it was not
unusual for the source to be detected in only one of the two
observations.

\begin{table}
  \begin{center}
    \begin{footnotesize}
      \begin{tabular}{l c c c r@{$\pm$}r r@{$\pm$}r r@{$\pm$}r r@{$\pm$}r r@{$\pm$}r r}
       Name & Type & Sem- & Band & \multicolumn{2}{c}{Distance} & \multicolumn{2}{c}I {Flux}
       & \multicolumn{2}{c}{RR Flux} & \multicolumn{2}{c}{LL Flux} & \multicolumn{2}{c}{Circ Pol}
       & S/N$^1$ \\
       &          &    ester    &         & \multicolumn{2}{c}{(arcsec)} & \multicolumn{2}{c}{($\mu$Jy)}
       & \multicolumn{2}{c}{($\mu$Jy)} & \multicolumn{2}{c}{($\mu$Jy)} & \multicolumn{2}{c}{(\%)} &  \\ \hline
      EQ Cet & AM & 15A & K & 0.78 & 0.30 & 96 & 30 & 111 & 55 & 0 & 55 & +100 & 70 & 11.2 \\
      Cas 1   & IP  & 13B & C & 1.87 & 2.50 & 21 & 10 & 0 & 33 & 117 & 32 & -100 & 39 & 9.7 \\
      FL Cet  & AM & 18A & X & 0.17 & 0.25 & 11 & 5 & 24 & 9 & 0 & 9 & +100 & 53 & 6.6 \\
      BS Tri   & AM & 15A & X & 0.07 & 1.00 & 57 & 9 & 49 & 16 & 0 & 15 & +100 & 45 & 23.3 \\
      EF Eri   & AM & 13B & X & 0.32 & 1.20 & 87 & 15 & 0 & 30 & 135 & 31 & -100 & 30 & 25.8 \\
      UZ For  & AM & 15A & C & 1.46 & 2.50 & 78 &  9 & 0 & 25 & 85 & 26 & -100 & 42 & 39.6 \\
      Tau 4    & AM? & 15A & X & 0.22 & 1.00 & 105 & 32 & 0 & 72 & 150 & 74 & -100 & 69 & 8.0 \\
      LW Cam & AM & 18A & X & 0.06 & 0.30 & 50 & 4 & 0 & 6 & 98 & 6 & -100 &  9 & $>$99.9 \\
      VV Pup  & AM & 13B & X & 0.04 & 0.95 & 79 & 14 & 82 & 25 & 0 & 24 & +100 & 42 & 20.4 \\
                   &       & 13B & K & 0.36 & 0.45 & 49 & 29 & 103 & 60 & 47 & 61 & +37 & 57 & 4.5 \\
      FR Lyn  & AM & 18A & X & 0.13 & 0.22 & 28 & 4 & 12 & 6 & 38 & 6 & -52 & 17 & 43.3 \\
      Hya 1   & AM & 18A & X & 0.04 & 0.23 &   6 & 6  & 0 & 8 & 29 & 8 & -100 & 39 &  7.3 \\
      HS0922+1333$^{\dagger}$ & AM & 18A & X & 0.03 & 0.22 & 8 & 5 & 5 & 7 & 3 & 7 & +25 & 100 & 4.5 \\
      WX LMi & AM & 15A & C & 0.06 & 0.38 & 73 & 12 & 49 & 17 & 26 & 19 & +31 & 34 & 22.8 \\
                  &       & 15A & X & 0.07 & 0.27 & 52 & 14 & 33 & 17 & 48 & 17 & -19 & 30 & 13.4 \\
      ST LMi$^*$  & AM & 13B & X & 0.08 & 0.70 & 153 & 12 & 0 & 30 & 221 & 31 & -100 & 20 & 95.2 \\
      AR UMa$^*$ & AM & 13B & C & 0.23 & 1.30 & 489 & 16 & 492 & 25 & 428 & 28 & +7 &  4 & $>$99.9 \\
                   &              & 13B & X & 0.19 & 0.78 & 432 & 14 & 439 & 24 & 396 & 24 & +5 &  4 & $>$99.9 \\
                   &              & 13B & K & 0.16 & 0.34 & 317 & 30 & 252 & 45 & 255 & 45 & -1 & 13 & 79.4 \\
      EU UMa & AM & 18A & X & 0.06 & 0.22 &  39 &   5 & 34 & 6 & 33 & 7 & +1 & 14 & 53.1 \\
      V1043 Cen & AM & 18A & X & 0.16 & 0.60 & 20 & 5 & 0 & 7 & 20 & 8 & -100 & 53 & 11.2 \\
      J1503-2207 & AM & 18A & X & 0.03 & 0.45 & 29 & 5 & 66 & 7 & 0 & 6 & +100 & 14 & 55.5 \\
      BM CrB  & AM & 13B & X & 0.06 & 0.81 & 43 & 15 & 0 & 24 & 83 & 24 & -100 & 41 & 10.3 \\
      MR Ser  & AM & 13B & C & 0.12 & 1.35 & 239 & 17 & 371 & 27 & 0 & 32 & +100 & 11 & $>$99.9 \\
                  &        & 13B & X & 0.06 & 0.78 & 116 & 15 & 221 & 24 & 0 & 24 & +100 & 15 & 65.2 \\
      MQ Dra & AM & 18A & X & 0.20 & 0.25 & 17 & 4 & 25 & 6 & 0 & 6 & +100 & 34 & 17.4 \\
      AP CrB  & AM & 18A & X & 0.11 & 0.30 & 24 & 4 & 18 & 6 & 20 & 6 & -5 & 22 & 27.0 \\
      Her 1    & AM & 15A & K & 1.27 & 0.42 & 48 & 17 & 106 & 33 & 0 & 34 & +100 & 45 & 14.2 \\
      V1007 Her & AM & 15A & X & 2.72 & 1.20 & 38 &  9 & 75 & 15 & 0 & 16 & +100 & 29 & 23.3 \\
      V1323 Her & IP & 15A & C & 2.16 & 1.90 & 43 &  9 & 80 & 12 & 0 & 11 & +100 & 20 & 32.2 \\
                       &     & 15A & X & 0.49 & 1.20 & 23 &  9 & 0 & 15 & 53 & 15 & -100 & 40 & 10.0 \\
      AM Her$^*$ & AM & 13B & C & 0.27 & 1.60 & 88 &  5 & 57 & 9 & 84 & 9 & -19 & 9 & $>$99.9 \\
           &                 & 13B & X & 0.13 & 0.92 & 192 & 12 & 176 & 17 & 171 & 17 & +1 & 7 & $>$99.9 \\
           &                 & 13B & K & 0.22 & 0.61 & 476 & 83 & 172 & 122 & 440 & 109 & -44 & 27 & 24.7 \\ 
      V603 Aql$^*$ & SH & 13B & C & 0.13 & 2.90 & 22 & 3 & 15 & 6 & 0 & 6 & +100 & 29 & 35.4 \\
           &                 & 13B & X & 0.16 & 1.70 & 32 &  7 & 35 & 12 & 0 & 12 & +100 & 59 & 13.2 \\
      V1432 Aql & AM & 18A & X & 0.04 & 0.26 & 15 & 5 & 21 & 7 & 0 & 7 & +100 & 47 & 9.3 \\
      J1955+0045 & AM & 18A & X & 0.09 & 0.28 & 79 & 5 & 70 & 6 & 73 & 7 & -2 & 6 & $>$99.9 \\
      QQ Vul  & AM & 13B & K & 0.10 & 0.50 & 92 & 39 & 0 & 56 & 134 & 58 & -100 & 60 & 36.9 \\
      AE Aqr$^*$  & IP & 15A & C & 0.01 & 1.39 & 5123 & 9 & 5048 & 20 & 5074 & 20 & 0 & 1 & $>$99.9 \\
           &                 & 15A & X & 0.01 & 0.88 & 5497 & 11 & 5462 & 24 & 5425 & 26 & 0 & 1 & $>$99.9 \\
           &                 & 15A & K & 0.03 & 0.40 & 8031 & 34 & 7924 & 54 & 7881 & 53 & 0 & 1 & $>$99.9 \\
      HU Aqr & AM & 18A & X & 0.25 & 0.23 & 44 & 13 & 16 & 13 & 64 & 13 & -60 & 23 & 18.4 \\
      V388 Peg & AM & 18A & X & 0.04 & 0.23 & 34 & 5 & 73 & 6 & 0 & 6 & +100 & 12 & 86.8
      \end{tabular}
    \end{footnotesize}
  \end{center}
  \caption{List of radio detected MCVs. \newline
    $^1$: The total S/N is the product of the S/N of the relative
    distance, flux, and polarized flux (see Section 3.1). \newline 
    $^*$: Previously known radio source. \newline
    $^{\dagger}$: S/N is the product of two detections.}
\end{table}

The detection rate from the shorter (2 and 5 minute) exposures in
Semesters 13B and 15A is about 16\% (19 of 122 stars). It increases to
about 33\% (14 of 42 stars) for the longer 20 minute exposures in
Semester 18A.  The detection rate at the C-, X-, and K-band
frequencies are 8\% (9 of 122 stars), 19\% (30 of 142 stars), and 6\%
(7 of 122 stars), respectively. The fraction of stars showing high
polarization in at least one observation or exposure is $\approx 73$\%
(24 of 33 stars). The number of stars showing little to no ($<10$\%)
polarization is about 24\% (8 of 33) and one star (V1043 Cen) shows
high polarization during the first detection and no polarization
during the second.

\section{Discussion}

\subsection{Unpolarized Emission}

Fourteen of the radio detections show unpolarized continuum emission
with a very high brightness temperature. Such emission is
characteristic of an incoherent radiative process in an optically
thick plasma. The radiation is believed to be mostly gyrosynchrotron
emission by mildly relativistic ($1 < \gamma < 5$) electrons gyrating
in a magnetic field. We believe that the source of the few tens of keV
to few MeV electrons are magnetic reconnection events in the region
between the two stars or near the surface of the late-type donor star.

\begin{figure}
  \includegraphics[width=6cm]{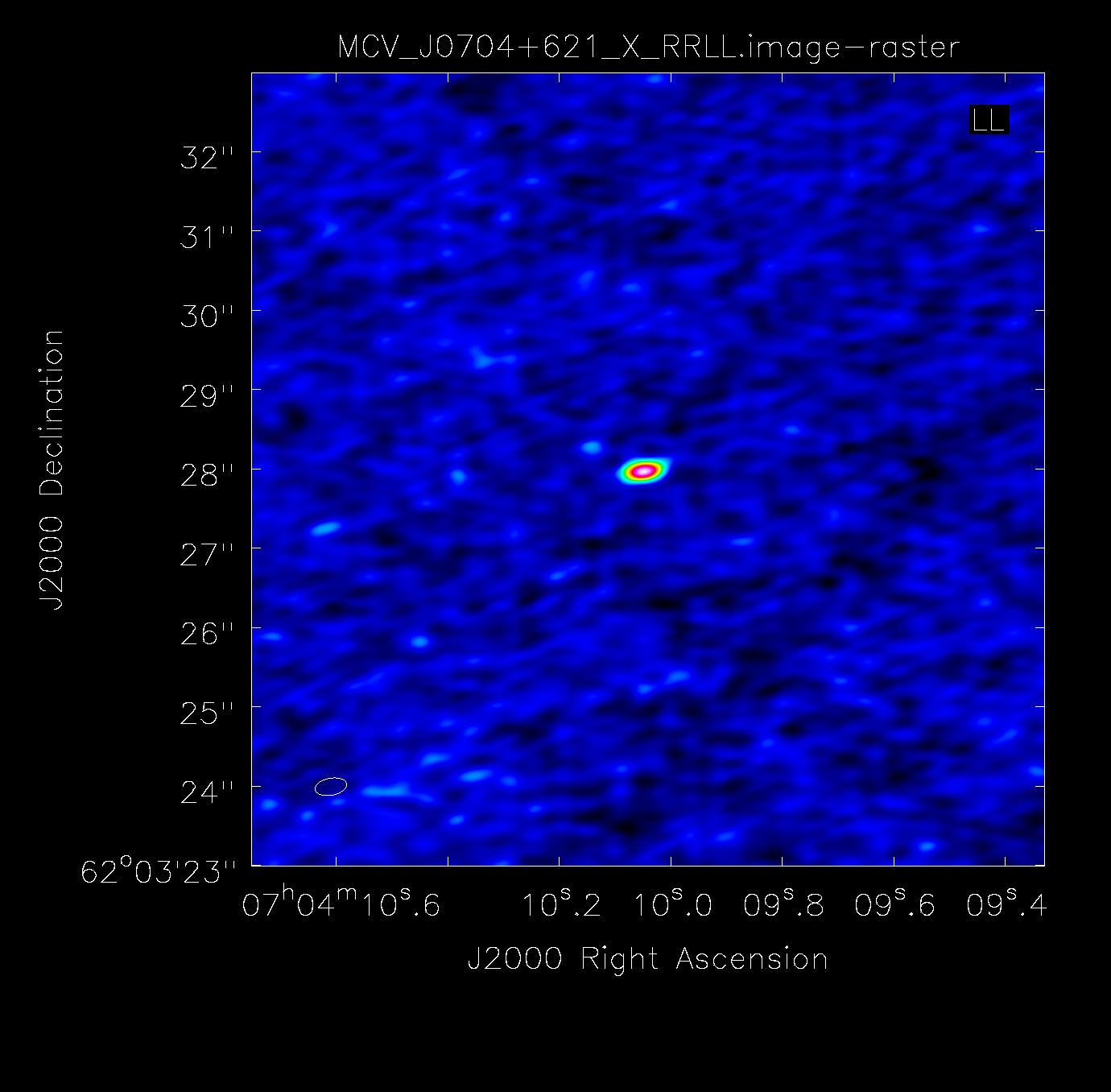}
  \includegraphics[width=6cm]{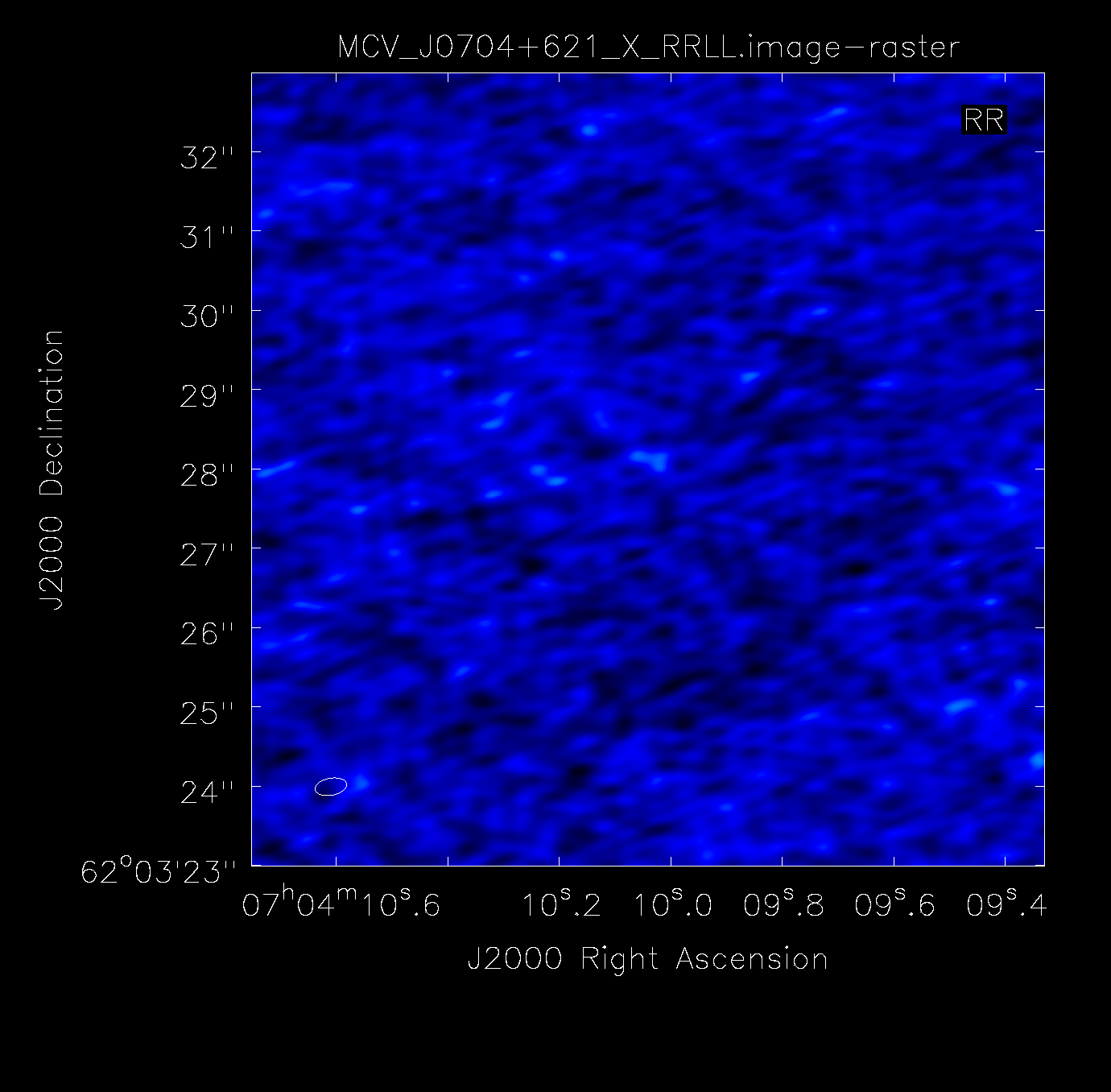}
  \caption{A ten minute exposure of the respective left and right
    circularly polarized X-band (10 GHz) images of the polar LW Cam
    showing the highly polarized radio emission.}
\end{figure}

\subsection{Polarized Emission}

A surprising result of this survey is that approximately three
quarters of the stars show highly polarized emission of short duration
($\sim$few minutes). An example of such emission is the left and right
polarized X-band images of the polar LW Cam (see Figure 1). The left
image shows an obvious point source, whereas the right does not.  High
circular polarization requires the presence of a strong magnetic field
(B $> 100$ G) and, therefore, limits the possible radiation mechanisms
to gyromagnetic and electron cyclotron maser emission. (see, e.g.,
\citet{dulk83, barr85, fuer86, benz89}. Gyromagnetic emission is often
divided into three energy regimes, non-relativistic ($\gamma = 1$)
cyclotron, mild-relativistic ($\approx 1 < \gamma < \approx 5$)
gyrosynchrotron, and relativistic ($\gamma > 5$) synchrotron emission,
where $\gamma = (1-(v/c)^2)^{-1/2}$. Each of these mechanisms has
problems reproducing the characteristics of the observed emission,
i.e., high polarization and short duration (see below). Synchrotron
radiation can be eliminated because its generates a broad continuous
spectrum and polarization of $< 40$\% \citep{bjor19}. Cyclotron and
gyrosynchrotron emission are also unlikely because very specific
conditions are needed to produce high polarization. Specifically, the
radiation must be emitted along the magnetic field in a relatively
homogeneous plasma \citep{barr85}. Any curvature of the magnetic field
and changes in plasma density and temperature will significantly
decrease the percentage of polarization. Therefore, the most likely
cause of the polarized emission is an electron cyclotron maser.
Theoretically, the characteristics of maser emission in a constant
magnetic field are high polarization ($>80$\%), short timescales
($<600$ s), and narrowband emission ($\Delta f/f < 0.01$, where f is
frequency; see. e.g. \citet{dulk83}).  The high polarization and short
duration are both seen in the roughly ten minute observations of LW
Cam and V603 Aql (see below). The absence of narrow band emission is
probably due to emission from a broad range of magnetic fields and
multiple emission regions.

One mechanism for producing the cyclotron maser is the loss-cone
instability. In this scenario, a plasma of relativistic electrons with
an isotropic velocity distribution is confined to a magnetic flux
tube.  The converging magnetic field creates a magnetic mirror causing
most of the electrons to be reflected, while the highest energy
electrons precipitate out. The loss cone anisotropy causes the index
of refraction of the plasma to be negative.  The magnetoionic modes of
the plasma are then amplified with the extraordinary (x) mode
typically growing much faster than the ordinary (o) mode. This
mechanism results in highly polarized radiation. \citet{dulk85} shows
that the growth rate of the x-mode for a typical loss-cone
distribution can be quite large ($\sim 10^7$ s$^{-1}$), which is
equivalent to an amplification length of $\approx 10$ m.

\begin{figure}
  \begin{center}
    \includegraphics[width=12cm]{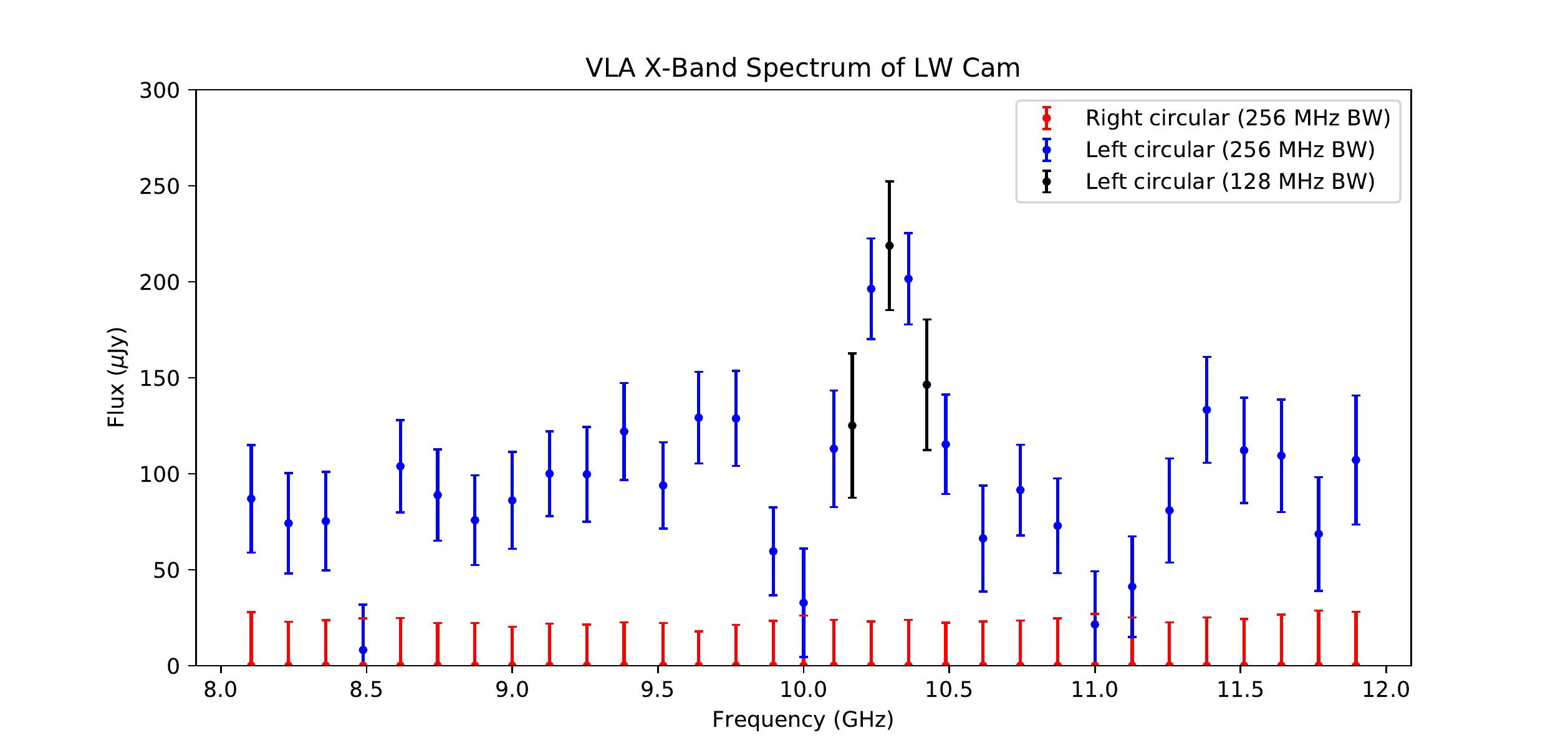}
  \end{center}
  \caption{A ten minute X-band (8--12 GHz) spectrum of the polar LW
    Cam.  The blue and red points are the left and right circular
    polarization channels, respectively. Each point is the average of
    two adjacent 128 MHz spectral windows in order to reduce the
    noise.  The spectrum suggests the presence of at least one
    narrowband ($< 300$ MHz) emission feature at about 10.295 GHz. The
    black points show the location of the peak.}
\end{figure}

Figure 2 is a ten minute X-band (8--12 GHz) spectrum of the polar LW
Cam. The blue and red points are the left and right polarization
channels, respectively. Each flux value is the average of two adjacent
128 MHz spectral windows (256 MHz bandwidth total), and their flux
centroids are within 0.3 arcseconds of the its Gaia position
\citep{gaia18}. This averaging of the data reduces the noise, but
smoothes the data, which may hide some narrow spectral
features. However, the spectrum does suggest the presence of at least
one narrowband ($<300$ MHz) emission feature at about 10.295 GHz,
which we identify as a pronounced electron maser event.  Assuming the
maser emission is at the fundamental harmonic, the ambient magnetic
field strength of the plasma is determined by the gyrofrequency:
f$_{ce} = 2.8$ B MHz, where B is in Gauss. For this particular event,
the magnetic field is $\approx$ 3.6 kG. We attribute the remaining
polarized emission to be the sum of many smaller maser events with
a range of magnetic field strengths.

Given the estimated strength of the magnetic field, there are two
possible source locations: in the accretion column and near the
surface of the donor star. First, suppose the emission is from the
accretion column and the ambient magnetic field is due to the white
dwarf. The magnetic field of the white dwarf is usually described by a
dipole with $B = 2 \times 10^7 r_9^{-3}$ G, where $r_9$ is the radius
of the WD in units of $10^9$ cm and the magnetic field of LW Cam is
$\approx 20$ MG \citep{ferr15}. The emission then originates at a
radius of about 18 WD radii, or about a third of the distance from the
WD to the inner Lagrangian point at 46 WD radii. The latter radius
assumes the WD and donor masses of 0.65 and 0.2 M$_{\odot}$,
respectively, and an orbital period of 97.3 minutes. For the relevant
equations, see equations 2.1b and 2.4c of \citet{warn95}. Within the
accretion column, the electron number density varies as
$n(r) = n_0 r_9^{-5/2}$ cm$^{-3}$, where $n_0 \sim 10^{16}$ cm$^{-3}$
at the base of the accretion column \citep{lamb73}. At 18 WD radii,
the electron number density and plasma frequency are $7 n_{12}$
cm$^{-3}$ and 24 GHz, resp., where $n_{12} = 10^{12}$ cm$^{-3}$ and
f$_{pe} = 9 n_{12}^{1/2}$ GHz.  A cyclotron maser requires f$_{ce} >$
f$_{pe}$. A condition that is not met within the accretion stream for
LW Cam at 8 GHz and is unlikely to be met for other polars at similar
frequencies. However, this argument does not exclude maser emission
from other low number density ($<10^{12}$ cm$^{-3}$) regions within
the WD magnetosphere.

In addition, the non-magnetic CV V603 Aql (Nova Aquilae 1918) also
shows highly polarized emission (see Table 1). The \citeauthor{ritt03}
catalog classifies this CV as a possible IP due to a prominent
photometric period, which explains its inclusion in our initial list
of targets. However, the photometric period is now associated with a
permanent superhump or precession of the accretion disk (see e.g.,
\citet{muka05}). If the WD has a magnetic field, it must be weak ($<1$
kG).  Otherwise, the Alfv\'en radius would be at least a few WD radii
(for a WD mass of 1.2 M$_{\odot}$ and an accretion rate of
$1.3 \times 10^{17}$ g s$^{-1}$) and would noticeably truncate the
inner edge of the accretion disk. The far ultraviolet study of V603
Aql by \citet{sion15} shows that this is not the case. They do not use
truncated accretion disk models to generate their synthetic
spectra. The accretion disk is also unlikely to be the site of the
polarized emission even when invoking a magneto-rotational instability
in the disk in order to increase angular momentum transport out of the
inner disk, because the magnetic field of the disk corona in such
models is $<1$ kG \citep{scep19}.

A more likely site of the polarized emission is in the lower corona of
the donor star where the plasma density and frequency are low enough
($n_e <10^{12}$ cm$^{-3}$ and $f_{pe} < 320$ MHz, resp.; \cite{dulk83,
  dulk85}) to allow cyclotron maser emission to escape.  The emission
site is probably near the footprints of one or more coronal loops with
the source of the high energy electrons being from magnetic
reconnection events in stellar flares. Although the mean coronal
magnetic field of the Sun is low ($<$ few G), because its fractional
surface coverage, or filling factor $f$, is very small ($<$1\%), the
magnetic fields in coronal loops can be $\sim$few kG
\citep{livi06}. In the case of dMe stars, their mean coronal magnetic
fields can be $\sim$few kG, because of their larger filling factors
($\sim$50\%). For the dMe stars AU Mic, AD Leo, and EV Lac;
\citet{saar93} estimates mean fields of 2.3 kG, 2.6 kG, and 3.6 KG,
respectively.  For this survey, the product of the measured field
strength in Table 2 and the estimated fillling factor, $fB$, can be
used as a rough estimate of the donor star's mean coronal magnetic
field. The data imply mean coronal fields of 1--4 kG, assuming a
filling factor of 50\%.  Higher field strengths are possible, since
the current upper bound is constrained by the observing frequency.

Table 2 lists the measured luminosities and magnetic field strengths
of the stars listed in Table 1. For stars with measurements at
multiple frequencies, only the largest luminosity and highest field
strength are listed. Columns 1-4 are the star's name, the CV subclass,
the orbital period in hours, and the star's distance in parsecs from
the Gaia Data Release 2. Column 5 is the flux value that is used to
calculate the radio luminosity in column 6. Column 7 is the estimated
magnetic field of the emitting region, which we attribute to maser
emission. We only list the measured magnetic field strengths for stars
having at least one measurement of high polarization. The field
strength is calculated using the central frequency and bandwidth of
the observation and therefore is only approximate. More accurate
estimates of the ambient magnetic field are possible for the brighter
stars via spectral analysis and may reveal the presence of multiple
emission features and hence, multiple simultaneous emission
regions. We intend to provide these results in a forthcoming paper.
The range of luminosities is shown as a histogram in Figure 3. Except
for AE Aqr, the highest luminosities are probably overestimates caused
by inaccurate Gaia distances.

\begin{figure}
  \begin{center}
    \includegraphics[width=12cm]{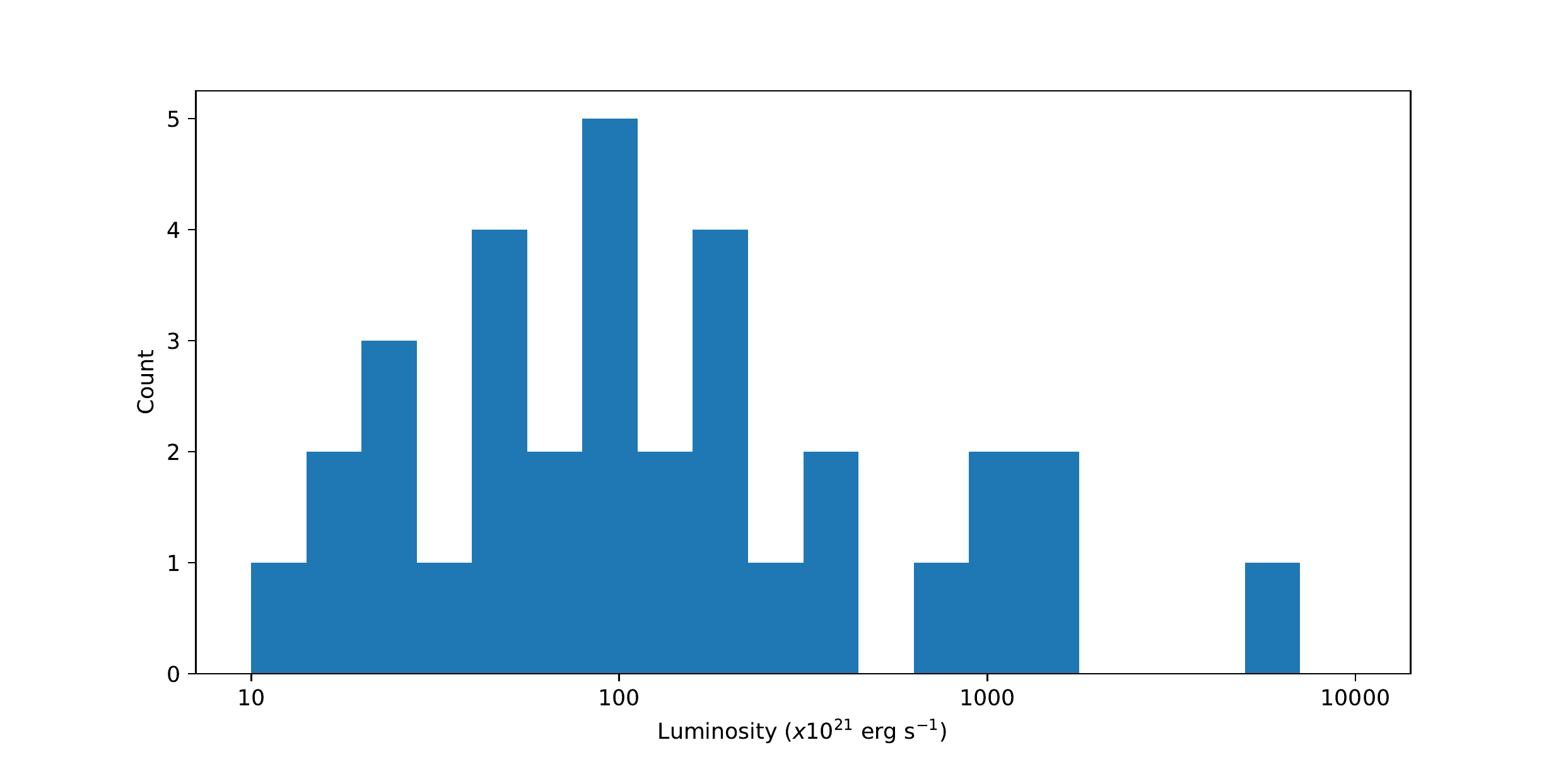}
    \caption{The number of MCVs versus luminosity. The sample is from
      Table 2.}
    \end{center}
\end{figure}

\begin{table}
  \begin{center}
    \begin{tabular}{l c c c c c r@{ $\pm$ }r}
      Name   & Type & Period & Distance & Flux & Luminosity$^*$ & \multicolumn{2}{c}{B-Field$^{\dagger}$} \\
                  &        &     (h)   &    (pc)  & ($\mu$Jy) & ($10^{24}$ erg s$^{-1}$) & \multicolumn{2}{c}{(Gauss)} \\ \hline
      EQ Cet  & AM & 1.5469 &  285   & 96 & 196 & 7530 & 365 \\
      Cas 1    & IP   & 3.9396 & 1608 & 21 & 292 & 1960 & 365 \\
      FL Cet   & AM & 1.4523 & 319 & 11 & 13 & 3563 & 730 \\
      BS Tri    & AM & 1.6045 & 273 & 57 & 46 & 3210 & 365 \\
      EF Eri    & AM & 1.3503 & 160 & 87 & 24 & 3210 & 365 \\
      UZ For  & AM & 2.1087 & 240 & 78 & 24 & 1960 & 365 \\
      Tau 4    & AM? & 1.50  ? & 210 & 105 & 50 & 3210 & 365 \\
      LW Cam & AM & 1.6211 & 542 & 50 & 175 & 3563 & 730 \\
      VV Pup   & AM & 1.6739 & 137 & 79 & 16 & 7530 & 365 \\
      FR Lyn   & AM & 1.8876 & 550 & 28 & 101 & \multicolumn{2}{c}{} \\
      Hya 1    & AM & 2.3966 & 441 & 6 & 14 & 3563 & 730 \\
      HS0922+1333 & AM & 4.0394 & 163 & 8 & 3 &  \multicolumn{2}{c}{} \\
      WX LMi  & AM & 2.7821 &  99 & 73 & 4 & \multicolumn{2}{c}{} \\
      ST LMi   & AM & 1.8981 & 113 & 153 & 21 & 3210 & 365 \\
      AR UMa & AM & 1.9320 & 101 & 489 & 27 & \multicolumn{2}{c}{} \\
      EU UMa  & AM & 1.5024 & 331 & 39 & 51 & \multicolumn{2}{c}{} \\
      V1043 Cen & AM & 4.1902 & 173 & 20 & 7 & 3563 & 730 \\
      J1503-2207 & AM & 2.2228 & 392 & 29 & 53 & 3563 & 730 \\
      BM CrB   & AM & 1.4040 & 419 & 43 & 81 & 3210 & 365 \\
      MR Ser   & AM & 1.8911 & 132 & 239 & 22 & 3210 & 365 \\
      MQ Dra  & AM & 4.3912 & 186 & 17 & 7 & 3563 & 730 \\
      AP CrB   & AM & 2.5310 & 209 & 24 & 13 & \multicolumn{2}{c}{} \\
      Her 1     & AM & 3.00  ? & 1030 & 48 & 1281 & 7530 & 365 \\
      V1007 Her & AM & 1.9988 & 462 & 38 & 87 & 3210 & 365 \\
      V1323 Her & IP & 4.4016 & 2244 & 43 & 1163 & 3210 & 365 \\
      AM Her  & AM & 3.0942 & 88 & 476 & 93 & \multicolumn{2}{c}{} \\
      V603 Aql & SH & 3.3168 & 313 & 32 & 34 & 3210 & 365 \\
      V1432 Aql & AM & 3.3656 & 462 & 15 & 38 & 3563 & 730 \\
      J1955+0045 & AM & 1.3932 & 171 & 79 & 28 & \multicolumn{2}{c}{} \\
      QQ Vul   & AM & 3.7084 & 317 & 92 & 233 & 7530 & 365 \\
      AE Aqr    & IP & 9.8797 & 91 & 8031 & 1673 & \multicolumn{2}{c}{} \\
      HU Aqr   & AM & 2.0836 & 192 & 44 & 19 & \multicolumn{2}{c}{} \\
      V388 Peg & AM & 3.3751 & 689 & 34 & 193 & 3563 & 730
    \end{tabular}
  \end{center}
  \caption{List of radio luminosity and donor magnetic field strength. \newline
    $*$ Luminosity assumes a 256 MHz bandwidth. \newline
    $\dagger$ The magnetic field assumes cyclotron maser emission from
    the corona of the donor star at the fundamental gyrofrequency, n
    =1.}
\end{table}

Under the assumption that the highly polarized radio emission is from
the donor star, these observations have important implications for
stellar dynamos in fully convective stars and for CV evolution.

\subsection{Cataclysmic Variable Evolution}

\begin{figure}
  \begin{center}
    \includegraphics[width=12cm]{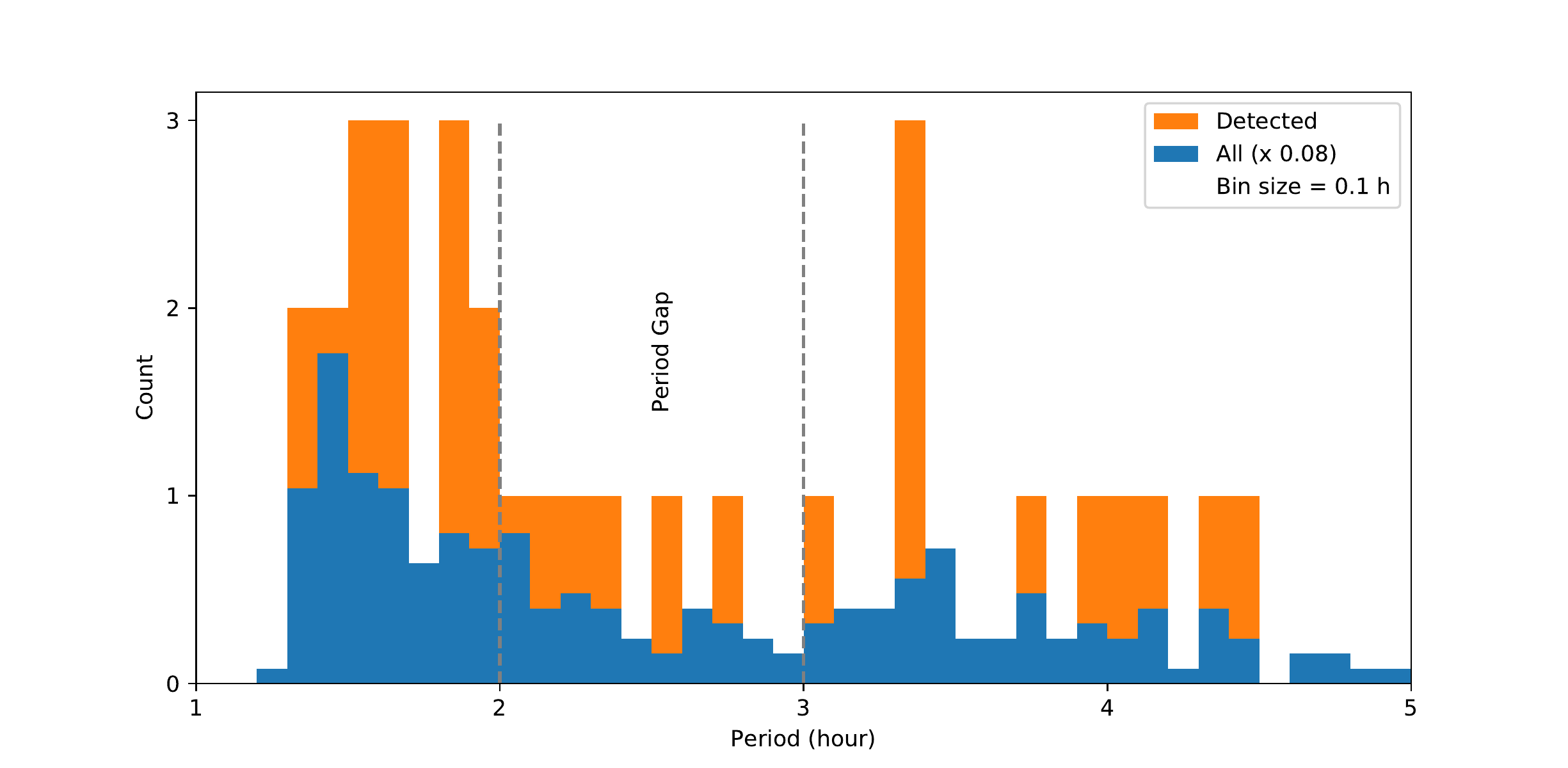}
    \caption{The number of detected MCVs (orange) and all MCVs (blue)
      versus orbital period. The detected sample is from Table 2. The
      vertical dashed lines delineate the period gap.}
  \end{center}
\end{figure}

\begin{figure}
  \begin{center}
    \includegraphics[width=12cm]{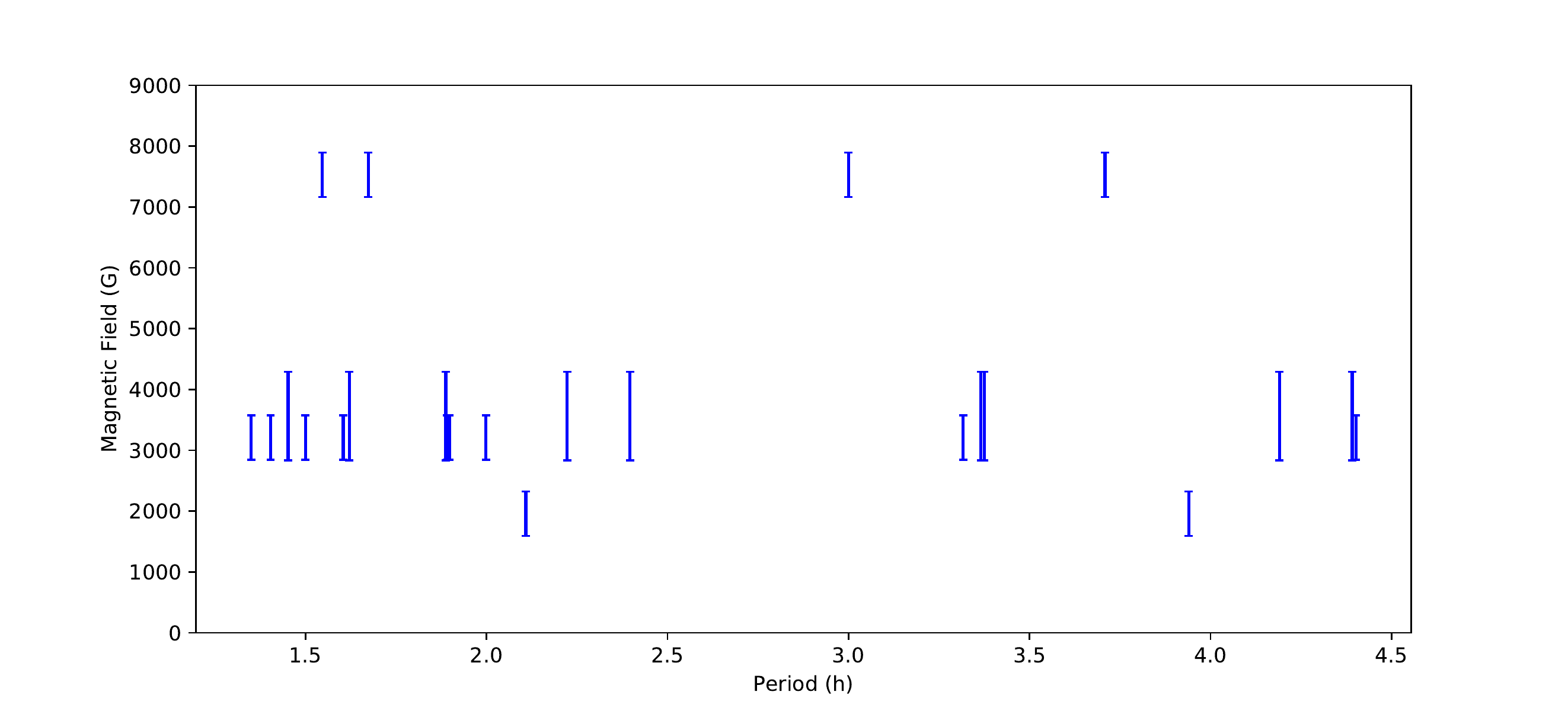}
    \caption{The approximate highest observed magnetic field strength
      of the electron cyclotron maser emission versus orbital
      period. The observing frequency is used to estimate the magnetic
      field strength of the emission region.}
  \end{center}
\end{figure}

The standard model of CV evolution is determined by angular moment
loss (AML; \citet{rapp82,spru83}). For P$_{orb} > 3$ hours, magnetic
braking is believed to dominate the AML, assuming the magnetic field
of the donor stars is $> 500$ G. The paucity of CVs with P$_{orb}$
between 2 and 3 hours was noted by \citet{whyt80} and is call the
``period gap''. \citet{robi81} suggested that the upper edge of the
gap coincides with the orbital period when the donor star becomes
fully convective. It has therefore been assumed that magnetic braking
is disrupted at P$_{orb} \approx 3$ hours. At this point, the donor
star shrinks and mass transfer ceases. Gravitational radiation then
becomes the dominating AML. At about two hours, the orbital radius
decreases sufficiently that the donor star overflows its Roche lobe
and mass transfer begins again.

The argument for fully convective stars having a weak magnetic field
is as follows. Donor stars above the period gap have a tachocline,
which is the interface between the radiative core and convective
envelope. The tachocline provides a seed magnetic field for the
magnetic dynamo in the convective envelope \citep{spie92}. Simulations
show that strong shearing occurs in the tachocline, causing the local
magnetic field to be amplified \citep{char14}, generating an
$\alpha-\Omega$ dynamo, named for the interplay between the cyclonic
eddies (the $\alpha$ effect) and the shearing of the field (the
$\Omega$ effect). Isolated main-sequence stars later than spectral
type M3--M3.5 (M $<3.5$ M$_{\odot}$) are predicted to be fully
convective and do not possess a tachocline. Therefore, it was
conjectured that they cannot sustain a strong dynamo, resulting in a
weak magnetic field (B $< 100$ G).

Observationally, fully convective stars do display all aspects of
stellar activity including optical variability, strong emission lines,
and ultraviolet, X-ray, and radio emission (see e.g.,
\citet{giam86,ster94,lins95,hodg95,flem95,delf98}). \citet{pevt03} has
shown that the X-ray activity of a late-type star is a good proxy for
the surface magnetic flux, and \citet{wrig16} have shown that the
level of X-ray emission as a function of stellar rotation period is
essentially the same for fully convective and partially convective
stars. Therefore, these stars must be capable of generating
significant magnetic fields. \citet{kuke99} studied dynamos in fully
convective pre-main-sequence stars using mean field theory. They found
that a second order effect, called the $\alpha^2$ dynamo, could be
excited for moderate rotation rates (P$_{spin} < 10$ d), giving rise
to steady, non-axisymmetric mean fields. Local fields as high as 10 kG
are generated in some simulations by \citet{brow08}.

Unlike previous studies that have indirectly inferred a strong
magnetic field for the donor star using stellar activity, our
observations provide direct measurements of the magnetic field in
coronal loops which are used to estimate the mean coronal magnetic
fields of the donor stars.  We conclude that all polars, and by
implication all CVs, have strong magnetic fields throughout their
evolution as a result of their rapid rotation. This can be seen in
Figure 4, where there is no decrease in the distribution function of
radio detected MCVs for P$_{orb} < 3$ hour. A two sample
Kolmogorov-Smirnov (K-S) test comparing the two orbital period
distributions functions of the radio detected MCVs and all MCVs gives
a statistic of 0.129 and a P-value of 0.71. The K-S test result
supports the conclusion that the orbital period distribution of radio
detected MCVs is the same as that of all MCVs. Figure 5 shows the
distribution of the approximate highest observed magnetic field
strength with orbital period. Although the data have a strong
observational bias because of the small number of detections and the
frequency bandwidth of the observations, there is no evidence of a
decrease in magnetic field strength with orbital period.  This
suggests that the magnetic fields of the donor stars for P$_{orb} < 3$
hours are just as strong as they are above 3 hours. This conclusion is
supported by recent work on dynamos of fully convective stars.

If the fully convective donor stars in CVs have strong magnetic
fields, then another mechanism must be the cause of the CV period
gap. One possibility is that there is a change in the structure of the
donor star's magnetic field. Dynamo theory has shown that partially
convective stars typically have dipolar or low-order multipolar
fields, while fully convective stars have high-order
multipoles. \citet{taam89} have shown low-order multipolar fields are
more efficient at AML than high-order multipolar fields.
\citet{garr18b} model CV evolution using the magnetic braking
prescription of \citet{garr18a} and show that it replicates the period
gap.  However, additional observations and analysis of the radio data
are needed to confirm \citeauthor{taam89}'s model.  We believe that
this explanation is consistent with the study of \citet{knig11} who
have shown that the optimal scaling factors for an empirical model of
AML above and below the period gap are: f$_{MB} = 0.66 \pm 0.05$ and
f$_{GR} = 2.47 \pm 0.22$, respectively, where f$_{MB} = $f$_{GR} = 1$
for the standard model of magnetic braking (MB) and gravitational
radiation (GR). A less efficient version of magnetic braking satisfies
the need for an additional AML mechanism to that of gravitational
radiation for P$_{orb} < 3$ hours.

\section{Conclusions}

Radio observations of magnetic cataclysmic variables provide a new
window for studying the radiation mechanisms and dynamics of these
interacting binaries. Under the assumption that the highly circularly
polarized emission is due to electron cyclotron maser emission from
coronal loops and the donor stars have a high magnetic filling factor,
we are able to directly estimate the mean coronal magnetic field
strengths to be about 1--4 kG. Although our sample size is limited (33
sources), a two sample K-S test supports our conclusion that the
distribution function of donor star magnetic fields is similar to that
all MCVs. This result implies that rapidly rotating (P$_{spin} < 10$
d), fully convective stars can sustain a strong magnetic dynamo, and
hence, strong coronal magnetic fields. It also suggests that magnetic
braking is important throughout the evolution of MCVs, and by
implication all CVs, and the change in AML across the fully convective
boundary at P$_{orb} \approx 3$ hours is the result of change in
magnetic field structure from low-order to high-order multipolar
fields as proposed by \citet{taam89}.

\section{Acknowledgements}

The National Radio Astronomy Observatory is a facility of the National
Science Foundation operated under cooperative agreement by Associated
Universities, Inc.

This work has made use of data from the European Space Agency (ESA)
mission {\it Gaia} (https://www.cosmos.esa.int/gaia), processed by the
{\it Gaia} Data Processing and Analysis Consortium (DPAC,
https://www.cosmos.esa.int/web/gaia/dpac/consortium). Funding for the
DPAC has been provided by national institutions, in particular the
institutions participating in the {\it Gaia} Multilateral Agreement.

The authors wish to thank the editor for her perseverance during this
project and to the two anonymous referees for their constructive
comments that greatly improved the final version of this paper.

\end{document}